\documentclass[prb,twocolumn,showpacs,superscriptaddress]{revtex4}
\usepackage{graphicx}
\usepackage{amsmath}
\usepackage{amsfonts}
\usepackage{epsfig} 
\usepackage{epstopdf}
\usepackage{color}
\usepackage{bm}
\usepackage{euscript,mathtools,amsthm,amsfonts,amssymb}
\usepackage[curve,knot,matrix,pdf]{xy}

\numberwithin{equation}{section}

\def\beq{\begin{equation}}
\def\eeq{\end{equation}}

\def\beqn{\begin{eqnarray}}
\def\eeqn{\end{eqnarray}}
\def\begeqar{\begin{eqnarray}}
\def\endeqar{\end{eqnarray}}

\begin{document}

\title{Pumping conductance, the intrinsic anomalous Hall effect, and statistics of topological invariants}
\author{Jan Dahlhaus}
\affiliation{Department of Physics, University of California,
Berkeley, CA 94720}
\author{Roni Ilan}
\affiliation{Department of Physics, University of California,
Berkeley, CA 94720}
\author{Daniel Freed\footnotemark[1]\footnotetext{These authors contributed primarily to the mathematical Appendix.}}
\affiliation{Department of Mathematics, University of Texas, Austin, TX 78712}
\author{Michael Freedman\footnotemark[1]}
\affiliation{Microsoft Research, Station Q, University of California, Santa Barbara, CA 93106, USA}
\author{Joel E. Moore}
\affiliation{Department of Physics, University of California,
Berkeley, CA 94720} \affiliation{Materials Sciences Division,
Lawrence Berkeley National Laboratory, Berkeley, CA 94720}
\date{\today}

\begin{abstract}
The pumping conductance of a disordered two-dimensional Chern insulator scales with increasing size and fixed disorder strength to sharp plateau transitions at well-defined energies between ordinary and quantum Hall insulators.  When the disorder strength is scaled to zero as system size increases, the ``metallic'' regime of fluctuating Chern numbers can extend over the whole band.  A simple argument leads to a sort of weighted equipartition of Chern number over minibands in a finite system with periodic boundary conditions: even though there must be strong fluctuations between disorder realizations, the mean Chern number at a given energy is determined by the {\it clean} Berry curvature distribution expected from the intrinsic anomalous Hall effect formula for metals.  This estimate is compared to numerical results using recently developed operator algebra methods, and indeed the dominant variation of average Chern number is explained by the intrinsic anomalous Hall effect.  A mathematical appendix provides more precise definitions and a model for the full distribution of Chern numbers.
\end{abstract}

\maketitle
\section{Introduction}

The problem of how conduction electrons move when the host metal is magnetic dates back at least to the measurement by Hall in 1879 of the effect that bears his name.  The term ``anomalous Hall effect'' refers to the observation that in many magnetic materials an electric field generates a transverse current density,
\beq
j_y = \sigma_{xy} E_x
\eeq
even in zero magentic field.  Current belief is that in most materials the primary effect on conduction electrons of the magnetism, even in a ferromagnet, comes from the modification of the electron band structure via spin-orbit coupling rather than from magnetic fields.  Recent years have seen a revival of interest including a great deal of progress on both the microscopic understanding of the effect~\cite{sundaramniu,sinitsynnomura} and its strength in various important materials~\cite{macdonaldiron,souzacobalt}; these accomplishments are summarized in a recent comprehensive review by Nagaosa {\it et al.}~\cite{ahereview}

One reason why this effect is particularly interesting is that it is believed to receive an ``intrinsic'' contribution from the Berry phase~\cite{berry} of the band structure, independent of the detailed nature of the disorder.  The Berry phase is a geometric property that probes how the Bloch states of the band electrons evolve as the crystal momentum changes.  As first discussed by Karplus and Luttinger~\cite{karplusluttinger}, the Berry phase appears in the semiclassical equations of motion of an electron wavepacket, and this was argued to explain the anomalous Hall effect.  Even for noninteracting electrons, the connection between the Berry phase and observed transport phenomena is quite subtle because scattering processes give additional contributions to the transverse conductivity $\sigma_{xy}$.  The goal of this paper is to use recent advances in the understanding of the Berry phase in {\it insulators}, specifically the integer quantum Hall effect
\beq
\sigma_{xy}  = {n e^2 \over h}, \quad n \in \mathbb{Z},
\label{IQHE}
\eeq
to give a different picture of the metallic anomalous Hall conductivity.

For most of this paper we will focus on two spatial dimensions, where (in the absence of spin-orbit coupling) the ultimate fate of a zero-temperature system is to be an insulator, either ordinary ($\sigma_{xy} = 0$) or quantum Hall (Eq.~\ref{IQHE}).  Our goal will be to show that the metallic anomalous Hall effect can still be understood by interpreting it as a disorder average over quantized insulators.  In each disorder realization, the insulator is characterized by a set of topological invariants determining $n$ in (Eq.~\ref{IQHE}), and $\sigma_{xy}$ is effectively quantized.  The average over disorder gives a non-quantized $\sigma_{xy}$, and we give an heuristic argument for why this averaged conductance should be determined in the weak-disorder limit by the intrinsic anomalous Hall formula. This approach gives a numerically accessible limit where the dominant, and possibly only, contribution to $\sigma_{xy}$ is the intrinsic contribution.  We then apply a recently developed operator algebra method~\cite{loringhastings} to compute the disorder-averaged conductance numerically for several models with nontrivial structure in $\sigma_{xy}$ as a function of electron density, which adds support to the theoretical picture.

We now introduce some basic notions and then describe the results more precisely.  The integer $n$ is well known to be quantized in two-dimensional insulators: it is determined by the sum of the Chern numbers or TKNN integers~\cite{tknn} of the occupied bands.  The Chern number is a topological invariant that is expressed as an integral over a band of the Berry curvature: defining the Berry connection or vector potential as
\beq
{\cal \bf A}^j = \langle u^j_k | - i \nabla | u^j_k \rangle,
\eeq
where $j$ is a band index and $k$ is crystal momentum,
the Berry curvature is
\beq
{\cal F}^j = \nabla_k \times {\cal \bf A}^j,
\eeq
and the Chern number of a band $j$ is
\beq
C_j = {1 \over 2 \pi} \int\,dk_x\,dk_y\,{\cal F}^j(k_x,k_y).
\label{Chern}
\eeq
Here the integral is over the Brillouin zone.  The total quantum Hall effect is then
\beq
\sigma_{xy} = {e^2 \over h} \left( \sum_{j\ {\rm occ}} C_j \right).,
\eeq
where in insulators each band is either fully occupied or completely empty.

The idea of the intrinsic anomalous Hall contribution is that, when a band is partially filled as in a metal, there remains something of the quantum Hall physics: there is a contribution to $\sigma_{xy}$ of the same form as the Chern number, except that the integral is only over the occupied states:
\beq
\sigma_{xy}^{\rm intrinsic} = {e^2 \over h} {1 \over 2 \pi} \int_{\rm occ}\,dk_x\,dk_y\,{\cal F}(k_x,k_y).
\label{intrinsic}
\eeq
(For conciseness, here and below we simplify the equations by assuming one partially filled band and ignore the quantized contribution from other bands that are completely filled.)  This integral can be transformed to a Fermi surface integral if desired~\cite{haldaneberry}.  We will review some of the standard arguments for this contribution in Section II.  The complications arise because of the disorder required to achieve standard metallic transport (as opposed to Bloch oscillations, where the electron distribution move periodically through the Brillouin zone with no relaxation).  With disorder, there are additional contributions to $\sigma_{xy}$ from impurity scattering (``skew-scattering'' and ``side-jump'' are two mechanisms that scale differently with the diagonal conductivity~\cite{ahereview}).  Disorder is particularly challenging to handle analytically in the two-dimensional case that we start with, as it is known that in a time-reversal breaking system (with or without spin-orbit coupling) the ultimate fate of noninteracting electrons at zero temperature and in the thermodynamic limit is always an Anderson insulator or a quantum Hall state.

For this reason we would like to find a fully quantum treatment of anomalous Hall transport that is non-perturbative in the disorder strength and hence does not use the semiclassical equations of motion.  It is known that if the disorder strength is kept constant as the system increases in size, eventually  ``quantum Hall plateau transitions'' develop~\cite{chalkercoddington,huckestein}.  (At very large disorder, all states become localized~\cite{yang&bhatt-1996} and there is only the ordinary insulator.)  The renormalization-group description of this process~\cite{khmelnitskiifloating,pruisken} leads to a two-parameter flow diagram, in which metallic conductivity appears only as an intermediate-coupling, nonuniversal effect.  The Hall conductance of a finite-size disordered system can be defined, under some assumptions, by an integral over periodic boundary conditions~\cite{niuthouless}, as reviewed in Section III.   In the simplest case of a transition from ordinary insulator to $n=1$ quantum Hall effect, virtually all realizations of disorder in a large system have $\sigma_{xy} = 0$ for filling less than some critical value $n_c$ and have $\sigma_{xy}=1$ above this filling.  An alternate case of Dirac fermions with spin-flip scattering where the Chern numbers move to band edges for strong scattering has been discussed~\cite{shengyuan}.  A metal does appear in some other symmetry classes: for example, in the symplectic ensemble describing 2D systems with spin-orbit coupling, there is a metallic phase between the ordinary and topological insulators (the latter is the quantum spin Hall phase~\cite{km2,zhangscience1,molenkampscience}).  This phase shows up as a region of fillings over which the Chern number continues to fluctuate as the system becomes larger.~\cite{essinmoore,loringhastings}

We will focus in this paper on approaching non-interacting metals in the unitary symmetry class (i.e., with broken time-reversal symmetry), essentially by scaling disorder to zero as system size grows.  The details of this approach are provided in Section III.  The basic idea is that, in addition to the plateau transition limit of strong disorder (more precisely, fixed disorder as system size grows), there is a weak-disorder limit where Chern numbers are defined but fluctuate strongly from realization to realization.  In other words, the Hall effect is strongly fluctuating between disorder realizations on a finite system.  However, the average remains meaningful and shows nontrivial structure very similar to that of the {\it metallic} anomalous Hall effect.  Section II reviews existing theories of the anomalous Hall effect and Section III introduces the theory of the Hall effect of a finite insulating system.  There is a natural conjecture for the disorder-averaged Hall effect of insulators, which we compare to numerical simulations of a variety of models in Section IV.  Section V discusses using the numerical results on the distribution of pumping conductance as input to macroscopic models of conduction in disordered magnetic materials.  An Appendix defines more precisely the mathematical question of how Chern numbers fluctuate between realizations and reviews some geometric considerations that enter into a simple model of the full probability distribution.

\section{Key features of the anomalous Hall effect}

The simplest approach to the anomalous Hall effect in metals starts from the semiclassical equations of motion for a wavepacket of Bloch states.  Remarkably, the AHE originates from a term in the semiclassical equations of motion that is neglected in almost all textbooks.  This term was first obtained by Karplus and Luttinger, but partly because their work took place before the understanding of Berry phases, their results were not universally accepted.  Consider the standard equations of motion~\cite{ashcroftmermin}
\beqn
\hbar {\bf \dot k} &=& e {\bf E} + e {\bf v} \times {\bf B} \cr
\hbar {\bf v} &=& \nabla_k \epsilon_n ({\bf k}) + \ldots.
\eeqn
where $\ldots$ indicate the often omitted Karplus-Luttinger term that we now explain.  As an electron wavepacket moves in $k$-space under the influence of an applied field, there are two contributions to its spatial velocity {\bf v}.   The first term, the group velocity, describes how the modified energy-momentum relation changes the velocity of the center of a wavepacket, which is an effect that would be present even for a point particle.  The Karplus-Luttinger contribution, which can be derived quite systematically~\cite{changniu,sundaramniu}, describes how a change in ${\bf k}$ induces a change in the real-space location because the Bloch states are changing: this change is
\beq
\hbar {\bf v}_{\rm KL} = - \hbar {\bf \dot k} \times {\bf \cal F}({\bf k}).
\eeq
Here ${\bf \cal F}$ is the Berry curvature defined in the introduction, and for two-dimensional systems there is a single nonzero component of the Berry curvature.  We will concentrate on the standard idealized limit where the time-reversal breaking effect on observables is not through an orbital $B$ field but rather through the Berry curvature ${\cal F}$.

Now the simple, if not entirely convincing, approach to the AHE in the DC limit is to compute the current from the anomalous velocity term using just the applied field ${\bf E}$ without considering the fields of impurities that normally act to produce a finite conductivity.  This could be justified by applying the Kubo formula with an order of limits that is not quite what takes place in an actual DC experiment: if the applied field ${\bf E}$ is extremely weak, then it takes a long time for the electrons to accelerate significantly from their ground-state distribution, and during this initial window the transverse current will indeed be given by
\beq
j_y = e \int\,{d^2 k \over (2 \pi)^2}\, v_y = {e^2 \over 2 \pi \hbar} E_x \left( {1 \over 2 \pi} \int\,d^2k\, {\cal F}_z \right),
\eeq
which is the same as Eq.~\ref{intrinsic}.   The integration is over the ground-state electron distribution, and time-reversal symmetry must be broken to give a nonzero answer since otherwise ${\cal F}({\bf k})$ is odd and $E({\bf k})$ even under ${\bf k} \rightarrow -{\bf k}$.  An objection, possibly first raised by Smit~\cite{smit}, is that, in the usual picture of DC transport in a metal, the system has reached a non-equilbrium steady state (at least to linear order, i.e., neglecting Joule heating) between the applied electric fields and electric fields from static impurities.  Then ${\dot \bf k}$ is zero and the anomalous velocity term does not contribute.

Hence the semiclassical approach might be useful at nonzero frequency but requires care at zero frequency, since a metal requires some level of disorder to have a finite DC conductivity.  In other words, it is not a priori clear why it should be allowed to regulate the divergent diagonal conductivity by introducing a relaxation time $\tau$, while assuming that the relaxation's effect on $\sigma_{xy}$ is controlled.  
Without disorder one would obtain Bloch oscillations, with no relaxation to equilibrium so that the usual assumption of the Kubo formula is invalid, and in the Bloch oscillation regime there would be no justification for just integrating over the ground state.  Diagrammatic treatments can be quite complicated and do not seem to be in clear agreement. There is substantial theoretical and experimental evidence~\cite{ahereview,PhysRevLett.103.087206} that the intrinsic formula (\ref{intrinsic}) does explain a large part of the DC conductivity in some magnetic metals.  For example, the orientation dependence of the conductivity tensor in anisotropic crystals is consistent with predictions from electronic structure calculations of the Berry curvature~\cite{souzacobalt}.




One reason for taking the approach in this paper is to see how the AHE formula can be understood while avoiding the criticisms of other approaches~\cite{gurevich,dyakonov}.  Part of this criticism in recent years grew out of consideration of the metallic spin Hall effect~\cite{sinova}, where at least for some disorder potentials it appears that the Berry-phase contribution is cancelled to 0~\cite{halperinmishchenko}.  The conclusion of the approach defined in the next section is that, for large finite samples, almost all of the variation of pumping conductance $\sigma_{xy}$ with energy is explained by the intrinsic contribution.  The only scope for a cancellation of the intrinsic part in the measured conductance is thus if the Fermi-surface effects, which are expected to depend on details of scattering, exactly cancel the scattering-independent intrinsic part.  The analytical picture and numerics presented here thus support the viewpoint of Ref.~\onlinecite{ahereview} that, even in more realistic 3D models (note that again treating finite systems a 3D model becomes equivalent to a large number of bands in a 2D model) at nonzero temperature in the thermodynamic limit, the intrinsic contribution can be dominant.  A way to state the resulting picture is that the pumping conductance computed below is more likely to be controlled by the intrinsic Hall effect contribution, as there is no Fermi surface so the other, scattering-specific contributions are minimized (or even zero).~\footnote{It should be noted that nothing is inconsistent with the mathematical statement that an integral of $F$ can be reformulated as a Fermi surface integral~\cite{haldaneberry}, since ultimately the only $F$ integrals computed in the following will be over filled bands}



\section{Pumping conductance and statistics of Chern number}

\subsection{Hall conductance in mesoscopic systems via pumping}

In the thermodynamic limit of a two-dimensional independent-electron system with broken time-reversal symmetry, there is no metallic region of energies.  Instead there are localized regions with occasional isolated energies supporting extended states of nonzero Chern number, and there is strong numerical evidence~\cite{huckestein,chalkercoddington} that the localization length diverges near these critical energies as
\beq
\xi \sim {1 \over (E-E_c)^\nu}.
\eeq
For a model system consisting of a tight-binding lattice Hamiltonian with ``topological bands'' having nonzero Chern number, to which on-site disorder is added, we expect at least two regimes.  For very strong disorder, all states are localized and have Chern number zero, which could be viewed as happening via ``floating'' of extended states of opposite Chern numbers until they annihiliate~\cite{yang&bhatt-1996} (note that the total Chern number of all the bands in a tight-binding model is always zero~\cite{ass}).  For moderate disorder, as the system size grows larger with disorder strength fixed, at least one sharp ``plateau transition'' energy becomes well-defined within a topological band, and the collapse of the transition width (i.e., the range in energy over which Chern numbers flucutate) with system size can be used as a measurement of the localization length exponent $\nu$.

Both of these limits have only insulating phases and cease to show strong fluctuations of the Chern number between realizations, except in a tiny range of energies around the plateau transitions.  The focus of this paper is on the fluctuation regime, which can be obtained in a large system by running the disorder to zero as the system size increases.  In order to explain the dominant effect observed in numerics in Section IV, we show how a simple theoretical argument leads to a conjecture for the average Chern number in the fluctuating regime that corresponds to the intrinsic anomalous Hall formula discussed in Section II.  In this section we give the theoretical motivation for the conjecture and discuss pumping versus Chern number.  Fluctuations of topological properties between disorder realizations are well known to appear in numerical approaches to metallic regimes, and continue to be studied actively~\cite{loringhastings,meyerrefael}, but we are not aware of a previous quantitative analysis of the low-disorder limit.

\begin{figure}
\includegraphics[width=1\columnwidth]{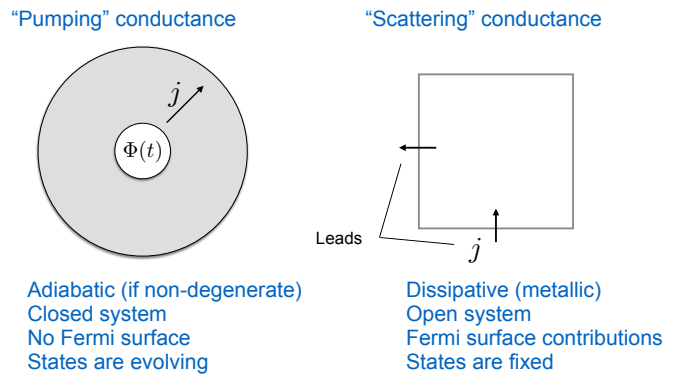}
\caption{\label{fig:pumping} {Comparison of pumping and scattering approaches to Hall conductance of a mesoscopic (i.e., finite) system.}}
\end{figure}

First, it is important to define what is meant by the zero-temperature conductance of a finite system.  The quantity we compute is the pumping conductance, as defined by Thouless et al.~\cite{tknn,niuthouless} to formalize ideas of Laughlin~\cite{laughliniqhe} shown in Fig.~\ref{fig:pumping}.  There are no leads and the transport is purely adiabatic: increasing a flux through the center of a Corbino disk, or through one great circle of a torus, leads to charge transport through the evolution of the ground state rather than through excitations.  Adiabatic transport is possible because we expect no level crossings for a 2D system in the unitary ensemble because a degeneracy has codimension 3 (i.e., requires tuning three parameters) and there are only two free parameters (the boundary fluxes through the two great circles of the torus).  

The Landauer conductance for an open finite system with leads corresponds to a different physical picture: the levels of the finite system are broadened and there is a nonzero diagonal conductivity, as opposed to the zero diagonal conductivity in the pumping picture.  The way these two pictures become equivalent in the thermodynamic limit is that it ceases to be possible to remain adiabatic at any fixed nonzero frequency, as the energy scale of avoided crossings goes to zero.  (Conversely, we expect the pumping conductance to be relevant to ``closed'' mesoscopic samples where the lead-induced linewidth is small compared to the level spacing.)  The quantization of transport in the pumping limit means that we can analyze $\sigma_{xy}$ in each disorder realization using the standard topological invariant, Chern number.  The disorder-averaged pumping conductance then becomes a statistical average of different Chern numbers in different realizations.


\subsection{The intrinsic anomalous Hall formula as a statement about average pumping conductance}

The simplest way to imagine how the Chern numbers might be distributed under the influence of disorder is as follows.  Return to a lattice system with a topological band, say with Chern number $+1$, and for simplicity let us assume that this band is the lowest in energy.  To study disorder, form a ``supercell'' of some large number $N$ of the original unit cells, and now allow on-site disorder (most kinds of local disorder that break all symmetries should behave similarly).  Before adding disorder, there would be $N$ connected bands obtained from the original band, and the only well-defined topological invariant is the total Chern number, still $+1$.  Now by the argument above, there will be no degeneracies in the new Brillouin zone, which is $N$ times smaller than the original unit cell but has $N$ times as many bands.  More precisely, the line crossings that are present in a supercell without disorder (since generically two surfaces will have linelike intersections) will no longer be present.  In the absence of intersections, each of the $N$ ``minibands'' will have its own Chern number, and all these Chern numbers will sum to $+1$.

At densities of $k/N$ electrons in the lowest band that fill an integer number $k$ of minibands in the Brillouin zone for the supercell, the pumping conductance will be given by the total Chern number of all filled minibands.  So we are led to consider the statistical problem of how minibands pick up Chern number in different disorder realizations.  The case of fixed disorder strength in the thermodynamic limit is discussed above: for disorder not too strong, there is a well-defined plateau transition at some density, and the metallic region of fluctuating Chern numbers around this transition decreases as a power-law in system size.

When disorder is very small, so that its only effect is to avoid band intersections, the metallic region of fluctuating Chern numbers extends over the whole band, and both positive and negative miniband Chern numbers are occasionally seen.  Consider two adjacent minibands.  If on average the weak disorder, in the process of eliminating band touchings, does not shift Chern number up or down in energy, then the mean Chern number assigned to a band can come only from the Berry curvature for the corresponding density range in the non-disordered problem.  In other words, the mean Chern number for each miniband will go to zero as $\sim 1/N$, and the proportionality constant is related to the {\it non-disordered} Berry curvature for the corresponding interval of density (or presumably energy in the weak-disorder limit).

Hence, writing $C_i$ for the Chern number of miniband $i$, this argument leads to the expression
\beqn
\langle \sigma_{xy}(E_F) \rangle &\equiv& {e^2 \over h} \lim_{N \rightarrow \infty \atop n_F/N = n(E_F)} \sum_{i = 1}^{n_F} \langle C_i \rangle \cr
&=& {e^2 \over h}\int_{E(k_x,k_y) < E_F}\,{F(k_x,k_y) \over 2 \pi} \,dk_x\,dk_y.
\label{conjecture}
\eeqn
So we obtain the conjecture that the disorder-averaged quantized pumping conductance of an ensemble of weakly disordered insulators is far from being quantized and actually given by the (non-disordered) intrinsic anomalous Hall effect expression.  The intrinsic anomalous Hall effect in metals is related to the averaged quantum Hall effect of insulators: this statement and its comparison to numerics are main results of the present work.  Construction of a possible model for the full distribution of the pumping conductance at each filling, not just its mean value, is described in the Appendix.

Clearly there are a number of assumptions in this argument that are difficult to justify: there is an assumption that the average over weak disorder does not move Chern number in a preferred direction in filling, for example.  
We will see in the following section, using recently developed numerical methods, that this prediction for the mean does correctly predict the dominant structure of the pumped conductance.  An interesting question to which we do not have a full answer is about the full distribution of miniband Chern numbers, beyond just the mean.  Both positive and negative integers appear, and a first approach is to construct a model for the Chern number based on (a) the conjecture for the mean above, and (b) random resolutions of the crossings that make the miniband Chern numbers independent and are almost statistically independent of each other (see Appendix).  Since the simplest experimental observable by far is the mean conductance, we will focus on numerical tests of the conjecture (\ref{conjecture}) here.





\section{Numerical methods and results}

\subsection{Approaches to efficient computation of Chern number for pumping conductance}

\label{sec:methods}
The first approach used to calculate Chern number was the conventional one of an integral over momenta (equivalently, over boundary phases~\cite{niuthouless} associated with phase-periodic boundary conditions on the torus).  In the supercell, we define the projection operator $P = |u_k^i \rangle \langle u_k^i |$ onto the subspace of miniband $i$.  Following Ref.~\cite{ass}, we can then compute the Chern number as the integral over momenta
\beq
C_i = {i \over 2 \pi} \int\,dk_x\,dk_y\,{\rm Tr}\, P (\partial_x P \partial_y P - \partial_y P \partial_x P).
\eeq
With a sufficiently fine mesh, this approach indeed gives quantization of subbands, but the fine mesh required means a considerable computational effort, limiting accessible system sizes to 10 by 10 or less.

To study larger system sizes and obtain a reasonable test of the conjecture regarding Chern numbers, we use the method of Hastings and Loring~\cite{hastingsloring,loringhastings}.  Let $P$ again be the projection operator onto some set of bands.  Then define two matrices $U$ and $V$ using the band-projected position operators in the two directions:
\beq
P \exp(i \Theta) P \sim \left(\begin{matrix}0&0\cr 0 &U \end{matrix} \right),\quad
P \exp(i \Phi) P \sim \left(\begin{matrix}0&0\cr 0 &V \end{matrix} \right)
\eeq

An integer (believed in general to equal the Chern number) can be defined that characterizes how far the $U$ and $V$ matrices, which are almost unitary and almost commuting, are from commuting:
\beq
{\rm Tr}\ \log (VUV^\dagger U^\dagger) = 2 \pi i m + r, \quad m \in \mathbb{Z}
\eeq
or alternately
\beq
\det(VUV^\dagger U^\dagger) = \exp(2 \pi i m + r), \quad m \in \mathbb{Z}
\eeq
where the definition of the determinant is by exponentiaing the trace of the logarithm as in the previous equation.  The integer $m$ can be nonzero once we note that there are many eigenvalues so that the total of their imaginary parts may be nonzero and equal to $2 \pi m$ even if each individual eigenvalue is .

A feature of this approach is that the method is constrained to return an integer for each band, even in the clean case where no Chern numbers for individual minibands are defined.  Hence we expect that at sufficiently small disorder the results cannot be trusted, and indeed at disorder values smaller than the ones in the following section, the Chern numbers become seemingly random small integers.  We have restricted ourself in the following to large enough disorder that the Chern number results are not distorted by this type of systematic error.

\subsection{Tight-binding models with nonzero Berry curvature}

\begin{figure}[tb] 
\centerline{\includegraphics[width=0.85\linewidth]{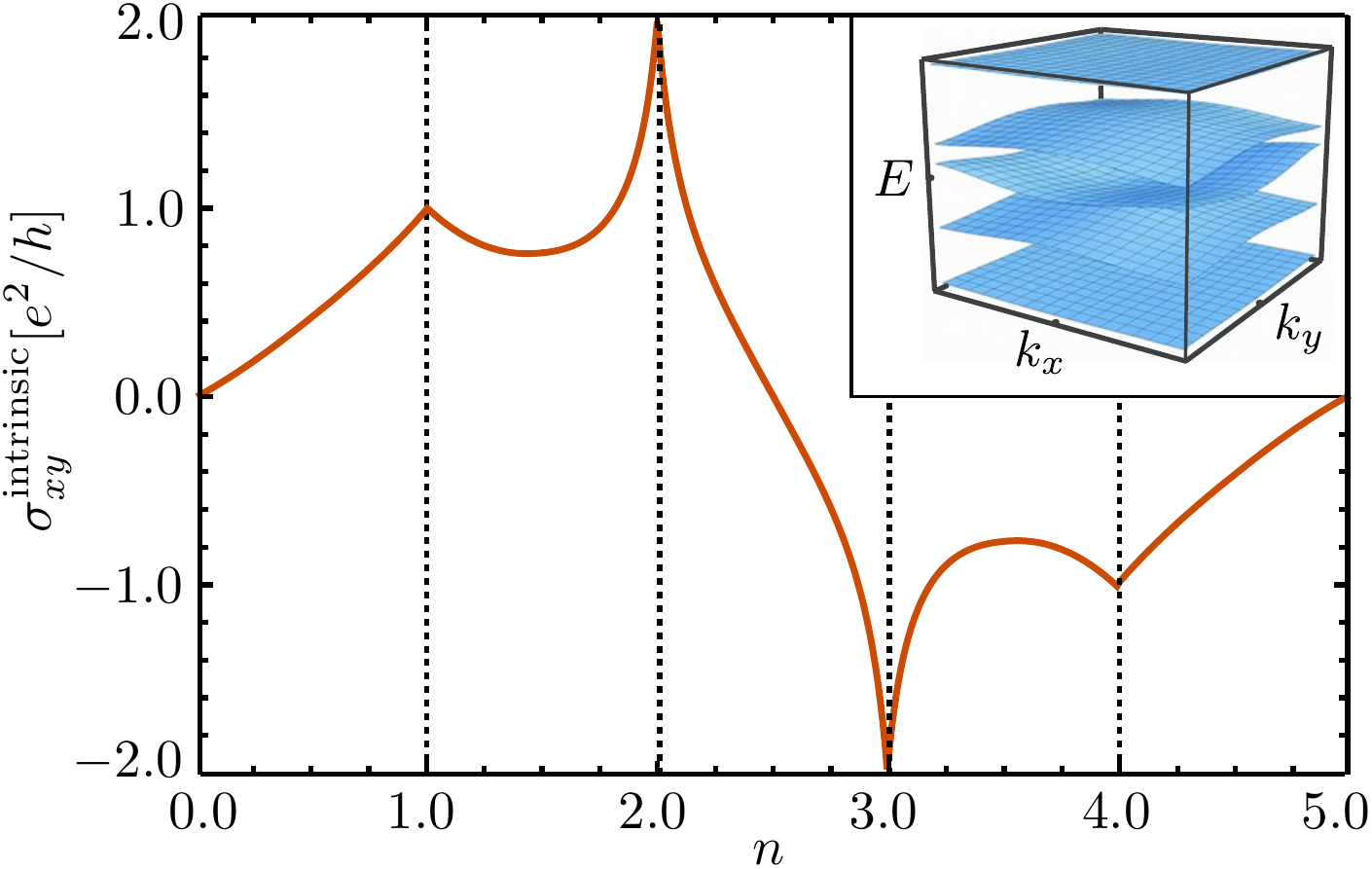}}
\caption{Intrinsic contribution to the Hall conductivity in the $1/5$ Hofstadter problem as a function of filling fraction, calculated from an integral over the Berry curvature, Eq. (\ref{QWZsigmaxy}). Dotted lines indicate when a band is completely filled and a new band begins. Inset: band structure of the model.
}
\label{fig:Fifthplots}
\end{figure}

The first model we  use to illustrate our findings is the typical Hofstadter problem on a square lattice,
\begin{align}
H_1&=\sum_{j,k}\left[ c_{j+1k}^\dag c_{jk}+ e^{-i j A_0} c_{jk+1}^\dag c_{jk}  + {\rm h.c.} \right],
\end{align}
in terms of creation operators $c_{jk}^\dag$ for particles on site ($j$,$k$) of the square lattice.
Electrons can hop from site to site while motion in $y$-direction comes hand in hand with a site-dependent phase $e^{-i j A_0}$ that reflects the presence of a uniform magnetic field perpendicular to the lattice plane. We concentrate on the special value $A_0=\Phi_0/5$ of the magnetic field, where each unit square contains one-fifth of a flux quantum and a unit cell of five sites can be defined. The electronic behaviour can then be described by a band structure with five bands as shown in the inset of Fig. \ref{fig:Fifthplots}. The figure itself shows the intrinsic contribution to the Hall conductivity as a function of the filling fraction of the bands, following the Berry curvature integral in Eq.  (\ref{QWZsigmaxy}). For filled bands (dotted lines), it equals the sum over the Chern numbers of the occupied bands. In this model, the bands carry Chern numbers $1,1,-4,1$ and $1$ respectively.

\begin{figure*}[bt] 
\centerline{\includegraphics[width=0.85\linewidth]{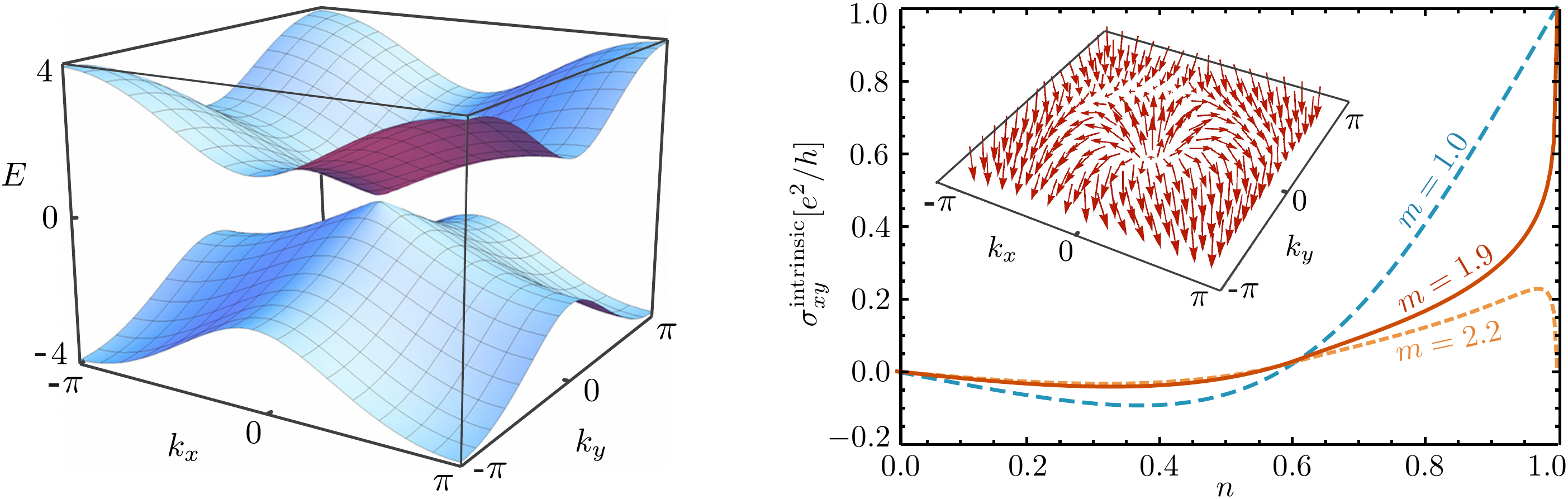}}
\caption{Left: band structure of the two band model in Eq.~(\ref{eq:2bandk}) for $m=1.9$. Right: intrinsic contribution to the Hall conductivity as a function of filling fraction of the lower band as calculated from an integral over the Berry curvature, Eq. (\ref{QWZsigmaxy}). As one can see at integer filling, $n=1.0$, the band carries a Chern number of one for $m<2.0$, while above the topological phase transition at $m=2.0$, the Chern number becomes zero. Inset: Spin quantization axis of the model in momentum space, illustrating the non-trivial winding that exists when the Chern number is one ($m=1.0$).
}
\label{fig:2bandplots}
\end{figure*}

The second model we consider is again a hopping model on a square lattice but this time with a spin degree of freedom. It is best defined for a clean system in momentum basis first, 
\begin{align}
H_2(\bm{k})=\bm{n}(\bm{k})\cdot \bm{\sigma}=
\begin{pmatrix}
\sin k_x\\
\sin k_y\\
m-\cos k_x- \cos k_y
\end{pmatrix}\cdot
\begin{pmatrix}
\sigma_x\\
\sigma_y\\
\sigma_z
\end{pmatrix},
\label{eq:2bandk}
\end{align}
with Pauli matrices $\sigma_i$ and momenta $k_x,k_y \in [-\pi,\pi)$ in the first Brillouin zone of the two-dimensional system. The two bands of the model, $E_\pm (\bm{k})=\pm |\bm{n}(\bm{k})|$, are plotted on the left in Fig. \ref{fig:2bandplots} for $m=1.9$.

 The corresponding hopping model reads
\begin{align}
H_2&=\frac{1}{2}\sum_{j,k,s}\left[   m \xi_s c_{jks}^\dag c_{jks}    - \xi_s [ c_{j+1ks}^\dag c_{jks}+c_{jk+1s}^\dag c_{jks} ] \right.\nonumber \\
&\hspace{52pt}\left. -i  c_{j+1k\bar{s}}^\dag c_{jks}+ \xi_s c_{jk+1\bar{s}}^\dag c_{jks}  +{\rm h.c.}   \right]
\end{align}
in terms of creation operators $c_{jks}^\dag$ for particles on site ($j$,$k$) of the square lattice with spin $s$. Here $\xi_\uparrow=1$ and $\xi_\downarrow=-1$ help to distinguish terms with different spin.

While the hopping model may seem complicated, the specific beauty and usefulness of this model comes with the fact that only two bands exist and that the Berry curvature can be expressed in terms of the normalized vector $\hat{\bm{n}}(\bm{k})=\bm{n}(\bm{k})/|\bm{n}(\bm{k})|$, yielding
\begin{align}
\sigma_{xy}^{\rm intrinsic}=-\frac{e^2}{h}\frac{1}{4\pi}\!\int\limits_{\rm occ}\!d k_x d k_y \!\left[\frac{\partial {\bm{\hat{n}}}(\bm{ k})}{\partial  k_x}\times \frac{\partial {\bm{\hat{n}}}(\bm{ k})}{\partial  k_y}\right]\!\cdot {\bm{\hat{n}}}(\bm{ k}),
\label{eq:WindingInvariant}
\end{align}
when only states in the lower band are occupied. The right part of Fig. \ref{fig:2bandplots} shows the intrinsic contribution as a function of filling for this clean hopping model. For a completely filled lower band, the intrinsic contribution $\sigma_{xy}^{\rm intrinsic}=\mathcal{I} e^2/h$ is determined by the Chern number 
\begin{align}
\mathcal{I}(m)=\left\{\begin{array}{cl}
\mbox{}{\rm sign}\,(m)&{\rm if}\;\;|m|<2,\\
0&{\rm if}\;\;|m|>2.
\end{array}\right.\label{QWZsigmaxy}
\end{align}
The vector $\hat{\bm{n}}(\bm{k})$ can be interpreted as the spin quantization axis for a given momentum and the Chern number becomes the winding number of this vector throughout the Brillouin zone, as illustrated in the inset of Fig. \ref{fig:2bandplots} for $m=1$, where $\mathcal{I}=1$.

\subsection{Numerical results with disorder}

Disorder is introduced in our models by adding a random potential energy on each site of the lattice to the Hamiltonian, 
\begin{align}
H_{1}^{\rm dis}&=H_{1}+\sum_{j,k} E_{jk} c_{jk}^\dag c_{jk},\\
H_{2}^{\rm dis}&=H_{2}+\sum_{j,k,s} E_{jks} c_{jks}^\dag c_{jks}.\\
\end{align}
We take these random energies from a gaussian distribution with zero mean and variance $U$. One can reduce undesired localization effects slightly by considering a different type of disorder distribution. We have tested this for a Wigner semicircle distribution - the results did not change much though. Our presentation will thus focus on Gaussian disorder.

To confirm our hypothesis numerically, we take disordered square samples of size $L\times L$ for either model and calculate the Chern numbers for the disorder bands following the methods mentioned in Sec. \ref{sec:methods}. We then plot the disorder average of the sum of Chern numbers of the first $l$ disorder bands against the ``filling'' $n=l/L^2$ and compare it to the intrinsic contribution to the Hall conductance of the clean case.

\begin{figure}[tb] 
\centerline{\includegraphics[width=0.8\linewidth]{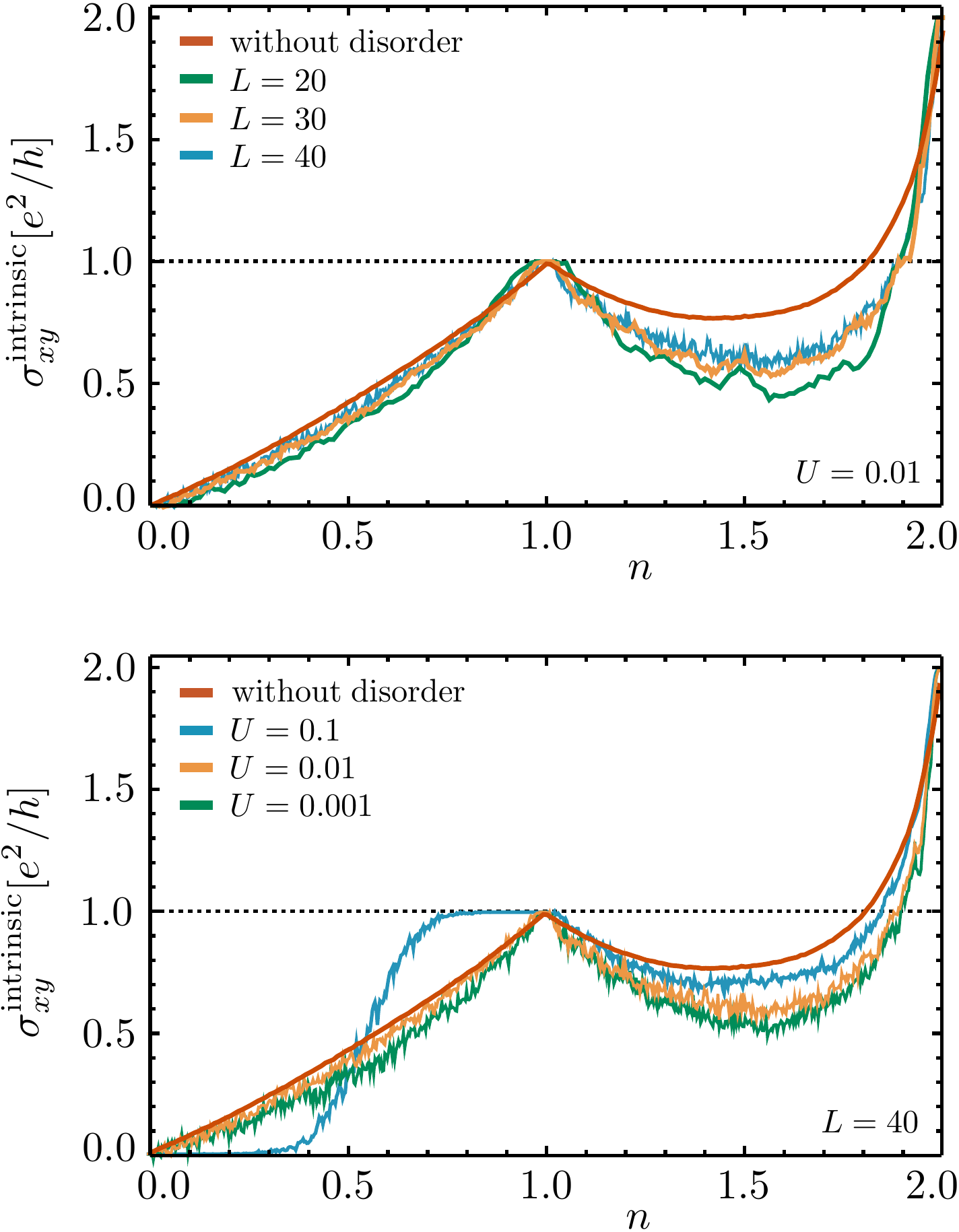}}
\caption{Disorder-averaged Chern sum of the first $l$ disorder bands plotted against "filling" $n=l/L^2$ for square samples of size $L\times L$ in the one fifth Hofstadter problem. Upper: sample size dependence for disorder strength $U=0.01$. Lower: disorder strength dependence for $L=40$. Averages were taken over more than $2500$ different combinations of random disorder configurations and random boundary phases. The red lines represent the intrinsic contribution to the Hall conductance in the clean case. 
}
\label{fig:FifthNumerics}
\end{figure}

Generally the numerical treatment has to be done carefully: a certain size is required so that sufficiently many disorder bands guarantee applicability of a statistical description. On the other hand Anderson localization sets in at a certain sample size for fixed disorder strength. This has to be avoided to maintain a metallic behaviour. Furthermore the numerical procedures only work reliably for large enough disorder strengths. For reasons of numerical efficiency, all results shown below have been obtained with the method of Hastings and Loring~\cite{hastingsloring} - agreement with the conventional method of integration over boundary fluxes has been established for smaller sample sizes though.

Fig. \ref{fig:FifthNumerics} shows our results for the disordered one fifth Hofstadter problem. We have focussed on the first two bands where all general trends can be observed. For the first band, the we find remarkable agreement between the disordered data and the intrinsic Hall conductivity in the clean case (red line) at sample sizes $L=20,30,40$ and $U=0.01$. One can see that the agreement becomes better with growing sample size. For the second band, the deviations between the clean and disordered cases are still a bit larger although the shape shows a clear resemblance. Naturally different bands may show a varying degree of convergence between the two cases. We suspect that larger samples (paired with smaller disorder) would bring the disordered curves closer to the red line - this is presently beyond the power of our numerical treatment. One can further see some bumps in the green ($L=20$) curve in the second band -  these are artifacts of too small sample sizes where the statistical treatment breaks down (the disorder band bandwidths are too large and the  bands overlap strongly). For even smaller samples (e.g. $L=10$), the curves deviate strongly and show additional undesired signatures.

The lower plot of Fig. \ref{fig:FifthNumerics} compares different disorder strengths. Disorder strengths $U=0.01$ and $U=0.001$ give practically the same results, the latter may already be effected slightly by numerical imprecisions. A deviation can be seen for $U=0.1$. In the lower band, localization sets in and flattens out the contribution of the outer disorder bands - around filling $0$ and $1$. The accumulation of Berry curvature is then concentrated on the band center. 
When investigating the behaviour of the second band, it may surprise to see better results for larger disorder. While the opposite behaviour would have been more reassuring, this doesn't constitute a problem per se, since there is an easy explanation: our conjecture is expected to hold for the small disorder - large system limit where the length of the system needs to be rescaled by $L \propto 1/U^2$ with decreasing disorder strength for curves to be comparable. Thus curves at fixed disorder are not required to show any clear relative behaviour when compared. An optimal treatment would show curves for different disorder strengths and fixed $U^2L={\rm const}$. Unfortunately the rather limited numerically accessible range of system sizes ($L\lesssim 40$) renders such an approach fruitless.

\begin{figure}[tb] 
\centerline{\includegraphics[width=0.8\linewidth]{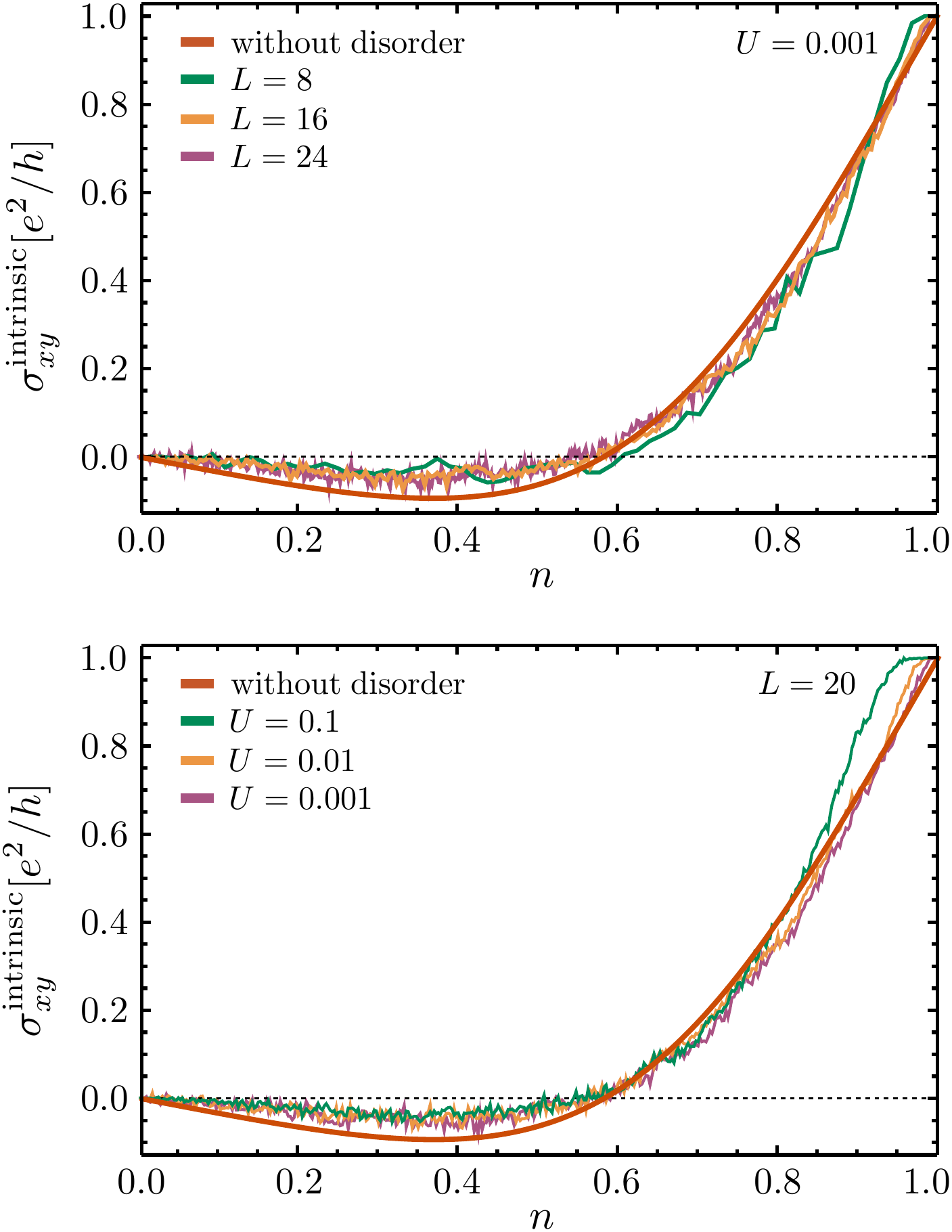}}
\caption{Disorder-averaged Chern sum of the first $l$ disorder bands plotted against "filling" $n=l/L^2$ for square samples of size $L\times L$ in the spinful two band problem at $m=1.0$. Upper: sample size dependence for disorder strength $U=0.001$. Lower: disorder strength dependence for $L=20$. Averages were taken over more than $10000$ different combinations of random disorder configurations and boundary phases. The red lines represent the intrinsic contribution to the Hall conductance in the clean case. 
}
\label{fig:2bandNumerics}
\end{figure}

\begin{figure}[tb] 
\centerline{\includegraphics[width=0.8\linewidth]{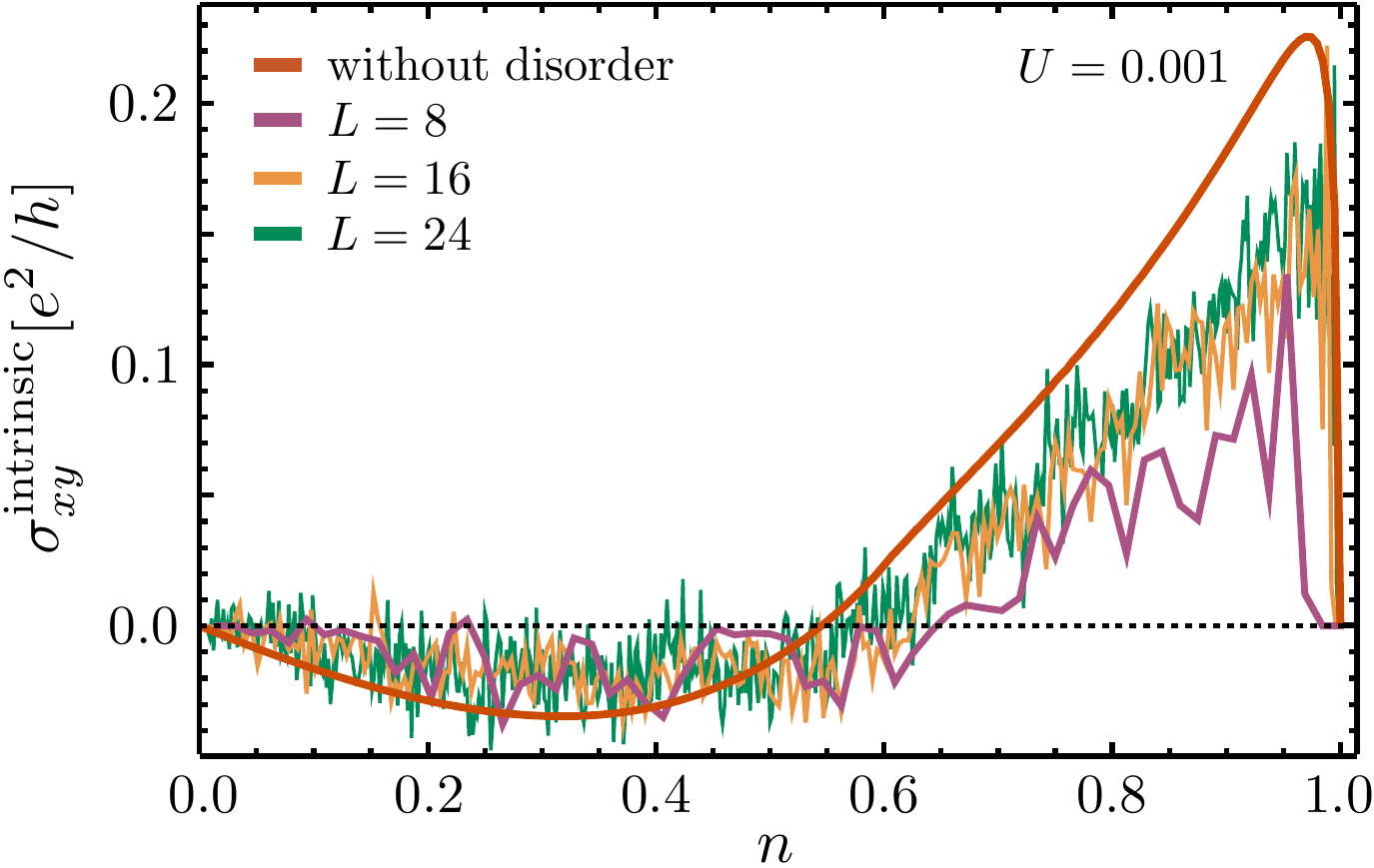}}
\caption{Disorder-averaged Chern sum of the first $l$ disorder bands plotted against "filling" $n=l/L^2$ for square samples of size $L\times L$ in the spinful two band problem at $m=2.2$. Upper: sample size dependence for disorder strength $U=0.001$. Lower: disorder strength dependence for $L=20$. Averages were taken over more than $10000$ different combinations of random disorder configurations and boundary phases. The red line represents the intrinsic contribution to the Hall conductance in the clean case. 
}
\label{fig:2bandNumericsTrivial}
\end{figure}

Let us now proceed to the second model (Eq. \ref{eq:2bandk}). Fig. \ref{fig:2bandNumerics} demonstrates the good agreement we obtain between the disordered data and the intrinsic Hall conductivity in the clean case (red line) at various sample sizes and disorder strentghs. We chose $m=1.0$ for the plots since then the gap between the bands is relatively large  and the bands carry Chern numbers $1$ and $-1$ respectively. Even though it is hard to see, larger samples provide better agreement (upper plot), just as lower disorder strength does. The green curve in the lower plot shows deviations due to localization of the outermost bands (around filling one), caused by the strong disorder $U=0.1$.
The two band model further allows us to explore the situation where the chern number of the bands is zero but where there are still Berry curvature effects.  In Fig. \ref{fig:2bandNumericsTrivial} we show data for $m=2.2$ where this is the case. Once again we find reasonable agreement which grows with system size. 

In summary our numerical results support our theoretical predictions qualitatively. A full quantitative agreement is out of reach due to the computational costs required to reach full convergence.

\section{Methods to model experiments}

In this section we discuss briefly how the purely quantum, $T=0$, finite-size results of this paper could be used as a basis to model experimental parameters under more realistic conditions.  At $T=0$, with fully coherent transport, there is no metal in two dimensions with broken time-reversal symmetry; rather, in the weak disorder limit discussed here, there would be strong sample-to-sample fluctuations between ordinary and quantum Hall insulators.  At nonzero temperature decoherence leads to a nonzero conductivity.  The problem of obtaining the effective conductivity tensor of disordered materials under time-reversal symmetry breaking is a widely studied one. These studies typically fall into two categories - one that studies a system with a continuously varying conductivity tensor, and one that refers to a system that is broken into regions with different conductivities and a sharp boundary between them, a so-called composite medium.  The result of the preceding sections, that a finite-size region of a disordered sample has a statistical distribution of pumping conductance in the weak-disorder limit, can serve as inputs to these larger-scale models.

The effective conductivity tensor of a system is defined as $$\langle j\rangle=\sigma_{e}\langle E\rangle$$ where $\langle j\rangle$ and $\langle E\rangle$ are the spatial (or volume ) averages of the local current density, $j(r)$, and electric field, E(r). These in turn are connected through the local relation $j(r)=\sigma(r)E(r)$, with $\sigma(r)$ the local conductivity tensor. The first approach to obtaining $\sigma_e$ is known as the effective medium approximation (EMA)~\cite{bergman1992physical,milton2002theory}, similar in spirit to a mean-field approximation, where a composite system is thought of as being made of grains with local conductivity $\sigma_l$, embedded in a uniform background material with conductivity $\sigma_{e}$, which is solved for self-consistently.  This method is most useful when considering a system made of a random mixture of two phases with scalar conductivities, where a closed form solution can be obtained for $\sigma_{e}$ as a function of the two local conductivities and the fraction of space they occupy. For non-scalar conductivities, which is the case of interest for systems with broken time reversal symmetry and a Hall-like response, this approximation allows for asymptotic solution at very high values of the magnetic field~\cite{bergman2000high}. 

Other methods focus on studying inhomogeneous quantum Hall systems near a plateau-plateau transition, taking a percolation theory approach where the picture of current networks forming at the boundary between quantum Hall puddles of different filling fraction is used to predict the scaling behavior of the Hall resistance.~\cite{simon1994explanation,ruzin1996nonuniversal} 
Two dimensional inhomogeneous media in a magnetic field can also be studied through using linear fractional transformations and duality relations~\cite{milton2002theory}. For example, in quantum Hall systems, fluctuations in the background potential cause the system to phase-separate into puddles of two quantum Hall phases with different Hall conductance. One way of applying linear fractional transformations for this scenario is by thinking of the local conductance as a complex number rather than a tensor, $\sigma_l=\sigma^l_{xx}+i\sigma^l_{xy}$ and mapping the system onto one with with zero magnetic field, where for a two phase composite exact solutions for the effective conductivity can be found. This way one can also derive a semi-circle relation between the longitudinal and the Hall conductance (another way is to map the system onto itself but flipping the direction of the magnetic field)~\cite{dykhne1971anomalous,PhysRevB.38.11296,milton2002theory,dykhne1994theory,ruzin1995universal}.

For more complicated scenarios where we expect the system to be composed of a random mixture of phases obtaining multiple values of the conductivity, a network model was formulated and numerically solved by Parish and Littlewood~\cite{parish2003non} originally as means to account for the non -saturating linear magnetoresistance observed in some inhomogeneous semiconductors. Within their model, the material is discretized into a random resistor network in a perpendicular magnetic field, where each unit is a four terminal disk. Four incoming and outgoing currents, $I_i$ are connected to four voltage differences $V_i$ via an impedance matrix $Z$, $$V_i=Z_{ij}I_j,$$ 
which is found by solving the Laplace equation with the appropriate boundary conditions. The parameters in $Z$ vary from unit to unit, generating the randomness of the network. The overall impedance of the network is the found numerically by using Kirchoff's law. The interesting behavior observed when mapping out the local current profile is that current loops form in the bulk, producing a Hall-like behavior with a Hall resistance linear in the magnetic field in the infinite network limit.

\section{Conclusions and future directions}

This paper demonstrates a close connection between disorder-averaged pumping conductance in weakly disordered mesoscopic systems and the instrinsic anomalous Hall effect in metals.  Only greatly simplified model band structures have been studied, and an obvious direction for future work is to see if the same strong correlation between pumping conductance and the intrinsic Hall effect formula is observed in realistic models of magnetic materials such as iron.  On the theoretical side, the Chern number is just one case of a topological invariant that is additive over bands, and the same question of how the topology is apportioned over minibands can arise for other symmetry classes and topological invariants.  A specific example is the symplectic case (i.e., a material with time-reversal symmetry and spin-orbit coupling), which in two dimensions has a $\mathbb{Z}_2$ invariant.  If the spin-orbit coupling conserves a $U(1)$ symmetry such as spin in the $z$-direction, then the problem will reduce to two copies of that studied in this paper.  For generic spin-orbit coupling, the statistical properties of $\mathbb{Z}_2$ invariants are an interesting question that requires a different theoretical approach than applied here.

The improvement of numerical methods~\cite{hastingsloring} to calculate topological invariants on a large system was an important enabling development for the present work, and we hope the present application will stimulate further methodological developments.  Another important theoretical development was the recent successful analysis of renormalization-group flow of conductivity in universality classes AIII and BDI in one dimension~\cite{altlandkamenev}, which have topology similar to that of the integer quantum Hall effect.  A similar analysis for the quantum Hall universality class should yield a limit of non-quantized $\sigma_{xy}$ obtained by reducing disorder with increasing system size as done here.  It would be quite desirable to develop a field-theoretic treatment of disorder averaging in the metallic regime that incorporates the microscopic information contained in the Berry curvature on momentum space in order to reproduce the observed dependence of $\sigma_{xy}$ on density.



The authors acknowledge useful conversations with A.~Altand and T.~Loring and support from a Rubicon fellowship (J.D.), AFOSR MURI (R.I.) and grant NSF DMR-1206515 (J.E.M.).  Parts of this work were completed at the Moorea Center for Advanced Studies.

\appendix
\section{Geometrical considerations and a model for statistical distribution of Chern number}

{\it Note}: This mathematical appendix was written by D. Freed and M. Freedman.  It provides a more precise mathematical definition of the main question defined in the main text and gives some geometrical background that is useful in understanding how Chern numbers are statistically distributed.  A simple model to capture these considerations is then defined.


As in the main text consider a two-dimensional planar system $H$ symmetric under a discrete lattice $Z\oplus Z$ of translations with Brillouin zone ($=$ momentum torus) $T\equiv T_\text{big}$.  Suppose, whimsically, we consider a larger ``super'' unit cell, $SC$, let us say consisting of $n^2$ original unit cells $C$. Correspondingly there is a small Brillouin zone $T_\text{small}$ and a covering map
\[\xymatrix{
Z_n\oplus Z_n \ar[r] & T_\text{big} \ar[d]^\pi \\
& T_\text{small}
}\]
where the $n^2$-sheets of the cover are permuted by the action of the covering group $G = \text{fine lattice} / \text{course lattice} \cong Z_n\oplus Z_n$.  Each character (homomorphism to $U(1)$) on the coarse lattice lifts to $n^2$ different characters on the fine lattice.

We consider a single low energy band of the system Hamiltonian $H$ giving a state (eigenvector of $H_k$) $\psi_k$ and an eigenvalue $e_k$, $k\in T_\text{big}$.  Thus $e$ is a function, $e : T_\text{big}\rightarrow\mathbb{R}$, and determines an $n^2$-sheeted ``multi-function'' $\tilde{e} := e\circ\pi^{-1} : T_\text{small}\rightarrow\mathbb{R}$.

The reason for considering the band as a multi-function on $T_\text{small}$ is to model disorder as:
\begin{enumerate}
    \item a passage from $C$ to super cell $SC$, to reflect reduced symmetry,
    and
    \item a resolution or ``avoidance'' of the crossings of $\tilde{e}$.
\end{enumerate}
This resolution produces precisely $n^2$-bands $e_i$, $i\leq 1\leq n^2$, $e_i : T_\text{small}\rightarrow\mathbb{R}$.  For topological reasons (the additivity of Chern classes - \cite{milnor}) the Chern number is conserved:
\begin{equation}\label{eq:conservation}
c_1(e) = \sum_{i=1}^{n^2} c_1(e_i).
\end{equation}
The ordered list of Chern numbers: $(c_1(e_1), c_1(e_2), \ldots, c_1(e_{n^2}))$, the ``Chern sequence'' $\vec{c}$, while deterministic in the precise choice of disorder, can also be regarded as a random variable on a measure space of possible disorder realizations (taken from disorder model in which parameters, e.g., ``strength of disorder'', have been fixed).

It looks exceedingly hard to solve a general instance of the deterministic problem so it may be of some value to propose a simple statistical model which captures key aspects of band resolution and requires as input only the most basic features of the band $e$ and the disorder model.  We write a Gibbs-type distribution for the Chern sequences $\vec{c}$ by assigning to each $\vec{c}$ a kind of mock energy $E(\vec{c})$, with $p(\vec{c}) = \frac{1}{Z} e^{-\beta E(\vec{c})}$, where the partition function $Z = \sum_c e^{-\beta E(c)}$.

Strangely, high $\beta$ (i.e., low temperature) corresponds to strong disorder, because as explained in the main text large disorder leads to the standard quantum Hall plateau transition at a fixed density and energy.  The fluctuating regime corresponding to high temperature is physically the metallic regime of weak disorder.

As input to the model we need two functions on $T_\text{big}^2$: the energy
$e : T_\text{big}^2\rightarrow\mathbb{R}$ introduced above and the Chern
density $c : T_\text{big}\rightarrow\mathbb{R}$.  The state $\psi_k$ is a
mapping $\psi _k: T_\text{big}^2\rightarrow BU(1)$ into the classifying space
consisting of rank $= 1$ Hermitian projectors $\psi_k\mapsto P_k :=
\left|\psi_k\right\rangle\left\langle\psi_k\right|$.  Concretely the Chern
form, $w$, pulls back to $T_\text{big}$ as: \begin{equation}\label{eq:form} w
:= \frac{1}{2\pi i}\operatorname{tr}\left(P\left(\frac{\partial P}{\partial
k_x}\frac{\partial P}{\partial k_y} - \frac{\partial P}{\partial
k_y}\frac{\partial P}{\partial k_x}\right)\right)dk_x\wedge dk_y,
\end{equation}
which specializes to the integrand of line ($20$) of this paper when the projectors act within $\mathbb{C}^2$.  Formula \ref{eq:form} also gives the $c_1$-integrand on all $BU(m)$ for general rank.  Besides $e$ and $c$ there is a parameter $\tau\in\mathbb{R}^+\cup 0$ that (roughly) measures influence of generic $\tilde{e}$-triple points: $\tau$ increases if there are more $k\in T_\text{big}^2$ with $e(k) = e(g_1(k)) = e(g_2(k))$ for $k\in T_\text{big}^2$ and $g_1$, $g_2\neq \text{id}$ are distinct elements of $G := Z_n\times Z_n$.  The final input is a choice of reciprocal temperature $\beta\approx$ disorder strength.  (A more precise understanding of this relationship would be desirable.)

Before describing the model we step back and give some abstract background
for the phenomena we propose to model.  The single band $\{\psi_k\}$ defines
a complex line bundle $E\rightarrow T_\text{big}^2$, i.e., the line
$E_k\equiv\mathbb{C}\psi_k$, the span of~$\psi _k$, lies over the point $k$.  $E$ pushes forward to a rank $n^2$ vector bundle $F\rightarrow T_\text{small}^2$ where the fiber $F_j$ over $j\in T_\text{small}^2$ is:
\[F_j = \bigoplus_{k\in\pi^{-1}(j)} E_k,\]
the direct sum being unordered.  Each summand is still labeled by its energy eigenvalue $e(k)$.

Topologically $F$ is isomorphic to a direct sum of $n^2$ line bundles.  We
give two reasons for this: the first abstract, the second concrete.

First,
the classifying space $BU\!\left(n^2\right)$ in dimensions $< 2n^2$ is homotopy equivalent to stable $BU$, which by Bott periodicity has:
\begin{equation}\label{eq:Bott}
\pi_{2(i-1)}(BU) \cong 0,\; \pi_{2i}(BU)\cong Z,\; i\geq 1.
\end{equation}
Thus any map $T_\text{big}\rightarrow BU\!\left(n^2\right)$, for dimensional reasons, factors through the $2$-skeleton of $BU\!\left(n^2\right)$ which it shares with $BU(1)$:
\[SK_2\!\left(BU\!\left(n^2\right)\right) =
SK_2\!\left(BU\!\left(1\right)\right) = \mathbb{C}\mathbb P^1\cong S^2.\]
This tells us $F$ actually has the form $(\text{line bundle})\oplus(\text{trivial line bundles})$.  This is \emph{more} than we want.  We want to arrive at a physically natural decomposition which in general will have many nontrivial factors.

The concrete picture is based on ``eigenvalue repulsion'' which states that
in families of Hermitian matrices collision of eigenvalues is a codimension
$3$ event.  Since $\operatorname{dimension}\!\left(T_\text{small}^2\right) =
2 < 3$, the multi-bands of $F$, $\tilde{e}$, will not collide if allowed,
through disorder, to become generic.  Then the eigenvectors associated to
$n^2$ \emph{disjoint} bands over $T_\text{small}$ define a decomposition into
line bundles $F\cong\oplus\,\text{line bundles}$.

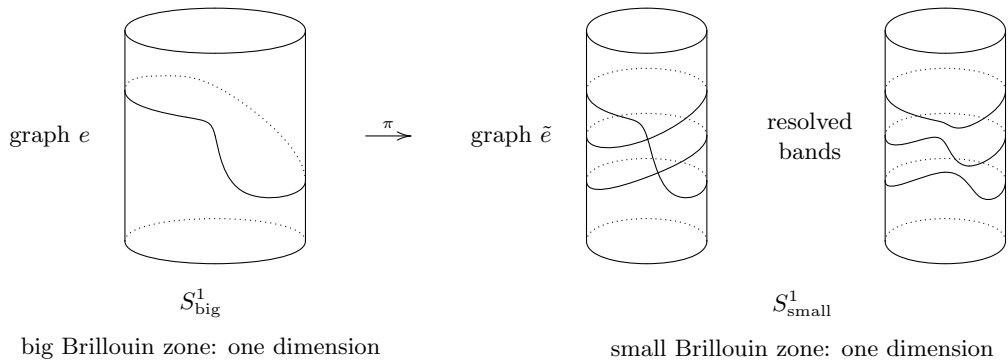
\begin{figure*}[hbpt]
\[\begin{xy}<4mm,0mm>:
(-3,0)="A",
(3,0)="B",
(-3,-7)="C",
(3,-7)="D",
(-3,-2)="A'",
(3,-5)="D'",
(-5.5,-3.5)*{\text{graph }e}
\ar@{-}@`{"A"+(0,-0.5),(0,-1),"B"+(0,-0.5)} "A";"B"
\ar@{-}@`{"A"+(0,0.5),(0,1),"B"+(0,0.5)} "A";"B"
\ar@{-} "A";"C"
\ar@{-} "B";"D"
\ar@{-}@`{"C"+(0,-0.5),(0,-8),"D"+(0,-0.5)} "C";"D"
\ar@{.}@`{"C"+(0,0.5),(0,-6),"D"+(0,0.5)} "C";"D"
\ar@{-}@`{"A'"+(0,-0.5),(-0.375,-3),(0,-3),(0.375,-6),"D'"+(0,-0.5)} "A'";"D'"
\ar@{.}@`{"A'"+(0,0.5),(-1.5,-1.5),(0,-1.5),"D'"+(0,1)} "A'";"D'"
\ar^-\pi (5,-3.5);(6.5,-3.5)
\end{xy}\hspace{8mm}
\begin{xy}<4mm,0mm>:
(-2,0)="A",
(2,0)="B",
(-2,-7)="C",
(2,-7)="D",
(-2,-2)="A'",
(2,-2)="B'",
(-2,-5)="C'",
(2,-5)="D'",
(-2,-3.5)="O1",
(2,-3.5)="O2",
(-4.5,-3.5)*{\text{graph }\tilde{e}}
\ar@{-}@`{"A"+(0,-0.5),(0,-1),"B"+(0,-0.5)} "A";"B"
\ar@{-}@`{"A"+(0,0.5),(0,1),"B"+(0,0.5)} "A";"B"
\ar@{-} "A";"C"
\ar@{-} "B";"D"
\ar@{-}@`{"C"+(0,-0.5),(0,-8),"D"+(0,-0.5)} "C";"D"
\ar@{.}@`{"C"+(0,0.5),(0,-6),"D"+(0,0.5)} "C";"D"
\ar@{-}@`{"A'"+(0,-0.5),(-0.5,-3),(0,-3),(0.5,-6),"D'"+(0,-0.5)} "A'";"D'"
\ar@{.}@`{"C'"+(0,0.5),(0,-4),"D'"+(0,0.5)} "C'";"D'"
\ar@{-}@`{"C'"+(0,-0.5),(0,-5),"O2"+(0,-0.5)} "C'";"O2"
\ar@{.}@`{"O1"+(0,0.5),(0,-2.5),"O2"+(0,0.5)} "O1";"O2"
\ar@{-}@`{"O1"+(0,-0.5),(0,-4),"B'"+(0,-0.5)} "O1";"B'"
\ar@{.}@`{"A'"+(0,0.5),(0,-1),"B'"+(0,0.5)} "A'";"B'"
\end{xy}\hspace{8mm}
\begin{xy}<4mm,0mm>:
(-2,0)="A",
(2,0)="B",
(-2,-7)="C",
(2,-7)="D",
(-2,-2)="A'",
(2,-2)="B'",
(-2,-5)="C'",
(2,-5)="D'",
(-2,-3.5)="O1",
(2,-3.5)="O2",
(-4.5,-3)*{\text{resolved}},
(-4.5,-4)*{\text{bands}}
\ar@{-}@`{"A"+(0,-0.5),(0,-1),"B"+(0,-0.5)} "A";"B"
\ar@{-}@`{"A"+(0,0.5),(0,1),"B"+(0,0.5)} "A";"B"
\ar@{-} "A";"C"
\ar@{-} "B";"D"
\ar@{-}@`{"C"+(0,-0.5),(0,-8),"D"+(0,-0.5)} "C";"D"
\ar@{.}@`{"C"+(0,0.5),(0,-6),"D"+(0,0.5)} "C";"D"
\ar@{-}@`{"C'"+(0,-0.5),(0.5,-4),(1,-6),"D'"+(0,-0.5)} "C'";"D'"
\ar@{.}@`{"C'"+(0,0.5),(0,-4),"D'"+(0,0.5)} "C'";"D'"
\ar@{-}@`{"O1"+(0,-0.5),(-0.5,-3.5),(0,-3.5),(0.5,-5),"O2"+(0,-0.5)} "O1";"O2"
\ar@{.}@`{"O1"+(0,0.5),(0,-2.5),"O2"+(0,0.5)} "O1";"O2"
\ar@{-}@`{"A'"+(0,-0.5),(-0.5,-3),(0,-3),(0.5,-3.5),"B'"+(0,-0.5)} "A'";"B'"
\ar@{.}@`{"A'"+(0,0.5),(0,-1),"B'"+(0,0.5)} "A'";"B'"
\end{xy}\]
\[\begin{xy}<4mm,0mm>:
(-9,0)="A"*{S_\text{big}^1},
"A"-(0,1.5)*{\text{big Brillouin zone: one dimension}},
(11,0)="B"*{S_\text{small}^1},
"B"-(0,1.5)*{\text{small Brillouin zone: one dimension}}
\end{xy}\]
\caption{One-dimensional illustration of the map from big to small Brillouin zone; these correspond to single unit cell and supercell in real space.}
\label{fig:1d}
\end{figure*}

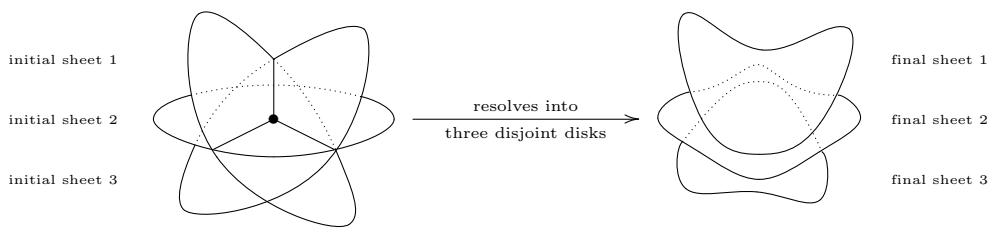
\begin{figure*}[hbpt]
\[\begin{xy}<4mm,0mm>:
(0,0)="O"*{\bullet},
(0,2)="A",
(-2,-1)="B",
(2,-1)="C",
(-2.5,3.5)="P1",
(2.5,-3.5)="P2",
(-4,0)="Q1",
(4,0)="Q2",
(-3,-3)="R1",
(3,3)="R2",
"P1"-(4.5,1.5)*{\text{\tiny initial sheet $1$}},
"Q1"-(3,0)*{\text{\tiny initial sheet $2$}},
"R1"-(4,-1)*{\text{\tiny initial sheet $3$}}
\ar@{-} "O";"A"
\ar@{-} "O";"B"
\ar@{-} "O";"C"
\ar@{-}@`{"P1"+(-0.48,-0.32),(-3.3,-2.2),"P2"+(-0.48,-0.32)} "P1";"P2"
\ar@{-}@`{"P1"+(0.48,0.32),(-1,3)} "P1";"A"
\ar@{-}@`{"P2"+(0.48,0.32),(2.6,-2.3)} "P2";"C"+(0.05,0)
\ar@{.}@/^2pt/ "A";"C"+(0.05,0)
\ar@{-}@`{"Q1"+(0,-0.5),(0,-2),"Q2"+(0,-0.5)} "Q1";"Q2"
\ar@{-}@`{"Q1"+(0,0.5),(-2.8,0.8)} "Q1";(-2.75,0.8)
\ar@{-}@`{"Q2"+(0,0.5),(2.8,0.8)} "Q2";(2.9,0.75)
\ar@{.}@`{(0,1.5)} (-2.75,0.8);(2.9,0.75)
\ar@{-}@`{"R1"+(0.4,-0.4),(2.8,-2.8),"R2"+(0.4,-0.4)} "R1";"R2"
\ar@{-}@`{"R1"+(-0.4,0.4)} "R1";(-2.625,-0.875)
\ar@{-}@`{"R2"+(-0.6,0.4)} "R2";"A"
\ar@{.}@/_3.5pt/ "A";(-2.625,-0.875)
\end{xy}\hspace{2.5mm}
\begin{xy}<5mm,0mm>:
\ar^{\text{resolves into}}_{\text{three disjoint disks}}(-3,0);(3,0)
\end{xy}\hspace{2.5mm}
\begin{xy}<4mm,0mm>:
(0,2)="A",
(-2,-1)="B",
(2,-1)="C",
(-2.5,3.5)="P1",
(2.5,-3.5)="P2",
(-4,0)="Q1",
(4,0)="Q2",
(-3,-3)="R1",
(3,3)="R2",
"R2"+(3,-1)*{\text{\tiny final sheet $1$}},
"Q2"+(2,0)*{\text{\tiny final sheet $2$}},
"P2"+(3.5,1.5)*{\text{\tiny final sheet $3$}}
\ar@{-}@`{"P1"+(0.48,0.32),"A"-(0,1),"R2"+(-0.6,0.4)} "P1";"R2"
\ar@{-}@`{"P1"+(-0.48,-0.32),"B",(0,-1.25),"C","R2"+(0.4,-0.4)} "P1";"R2"
\ar@{-}@`{(-2.8,0.8),"Q1","B",(0,-2.5),"C","Q2",(2.8,0.8)} (-2.35,0.9);(2.65,0.8)
\ar@{.}@`{(-1,1),"A"+(0,0.75),(1,0.75)} (-2.35,0.9);(2.65,0.8)
\ar@{-}@`{(-2.75,-0.9),"R1",(0,-2),"P2",(2.3,-1.2)} (-2.55,-0.75);"C"+(0.1,-0.1)
\ar@{.}@`{(-1,1.25),(0,1.25),(1,1.25)} (-2.55,-0.75);"C"+(0.1,-0.1)
\end{xy}\]
\caption{Resolution of triple points into disjoint disks under weak disorder.}
\label{fig:triple_point}
\end{figure*}

Let us look first at the topological picture of resolving a multi-band in one dimension, for, say, $n = 3$ (see Figure \ref{fig:1d}).
In two dimensions the situation is similar but there are now generically triple points of bands to consider.  Looking down from the top, say, at the graph of the multi-function $\tilde{e} : T_\text{small}^2\rightarrow\mathbb{R}$ we see triple points as in Figure \ref{fig:triple_point}.

The resolution of triple points together with the previously shown one
dimension resolution (Figure \ref{fig:1d}), but now in $1$-parameter families
along double point arcs and double point circles, occurs when the Hamiltonian
$H_k$ is perturbed to avoid eigenvalue collision.  The result is some
explicit way of decomposing $F$ as a sum of line bundles.  By
\eqref{eq:conservation} the total Chern number is conserved, and this is all we know rigorously.  But we now present a detailed heuristic for how to understand this conservation law as locally implemented by the resolution of double and triple points.  This perspective, although heuristic, suggests a way to model the ``diffusion'' of curvature between nearby small bands.

In mathematics, symmetry often destroys genericity and so it is here: the $G = Z_n\times Z_n$ symmetry of the cover $T_\text{big}^2\rightarrow T_\text{small}^2$ forces a non-generic eigenvalue distribution on $F\rightarrow T_\text{small}^2$.  Disorder breaks this symmetry and  perturbs the $n^2$ bands to be disjoint, much as drawn in Figure \ref{fig:triple_point}.  The \emph{weaker} the disorder the closer the classifying map comes to the co-dimension $= 3$ eigenvalue-collision locus $X$, so the more violently the eigenspaces must move in following the resolution which carries them between orthogonal eigendirections.  If the resolution is on a (momentum space) length scale $l$, the velocities of the eigenvectors scale as $l^{-1}$.  This leads to our interpretation of effective temperature in the model.

A final preliminary to defining the model is to understand a conservation law pertinent to the resolution of a single double circle between two sheets of $\tilde{e}$.  Consider an angular region $A\subset T_\text{small}^2$ around the double curve, and for simplicity assume no other double curves meet $A$.  Imagine the disorder perturbs $A$ relative its boundary $\partial A$.  Before the perturbation interior $A$ meets $X$ (actually along a circle but this is not relevant).  The universal first Chern form $w$ for the two bands integrates to $\pm 1$ respectively on the linking $2$-sphere to $X$; it is the Bott generator.

The perturbation to $\widetilde{A}$ relative $\partial A$ will generically contribute oppositely: $\pm\int_{\widetilde{A}} w$ to the two bands.  The contribution is not generally integral since $\partial\widetilde{A} = \partial A\neq\emptyset$.  Thus when bands are resolved we expect bits of curvature to be added to one band and subtracted from an adjacent band.  Triple points make the picture still more complicated by sharing curvature between triples of adjacent bands.  We model this effect by introducing a correlation coefficient, $\tau$.  We now define the model.

Over $T_\text{small}^2$ we define the pre-resolution bands $e_i^\prime$, $1\leq i\leq n^2$, as $e_i^\prime(j) = i^\text{th}$ largest value among $e(k)$ where $\pi(k) = j$.  The $e_i^\prime$ are piecewise smooth and we define $a_i = \int_{e_i^\prime} w$.  Note that $\{a_i \mid 1\leq i\leq n^2\}$ can be computed from the functions $e$ and $c$.  Now consider all possible ``double point'' corrections $d_i$, $1\leq i<n^2$, so that the following numbers are integers:
\beqn
\vec{c} &=& c_1 = a_1 + d_1, c_2 = a_2 - d_1 + d_2, c_3 = a_3 - d_2 + d_3, \ldots, \cr
&&c_{n^2} = a_{n^2} - d_{n^2 - 1}.
\eeqn
Clearly $d_1\in Z - a$, $d_2\in Z - a_2 + d_1, \ldots$  The $c$'s represent the Chern classes on the resolved bands $\overline{e_i}$.

The ``mock energy'' we use is:
\[E\!\left(\vec{c}\right) = \sum_i^{n^2 - 1} \left(\left|d_i\right| + \tau\left|d_i - d_{i+1}\right|\right),\]
where by convention $d_{n^2} = 0$.

\noindent\emph{Remarks.} The philosophy is that the resolution as a map to the classifying space
\[\bar{e} := \coprod_{i=1}^{n^2} e_i : \coprod_{i=1}^{n^2} \left(T_\text{small}^2\right)_i \rightarrow \prod_{i=1}^{n^2} BU(1) \subset BU\!\left(n^2\right)\]
has a geometrically natural energy, $\int d\,\text{area}\left|\nabla\bar{e}\right|^2$, of which the portion integrated away from the multiple points is fixed by the data $e$ and $c$.  For the integral near the double points we use the Chern curvature density as a surrogate for energy.  There is a Wirtinger-type inequality:
\begin{equation}\label{eq:Wirtinger}
\left|\text{Chern curvature }\bar{e}\right|\leq\text{const.}\left|\nabla\bar{e}\right|^2,
\end{equation}
and we are using the left-hand side to estimate the right-hand side.  This is likely a better approximation the higher the disorder.  Disorder sets the length scale (in momentum space) over which $\bar{e}$ can relax toward a map using only the topologically mandated energy.  Similarly, the longer this length scale, the more the perturbed system explores its phase space before choosing the curvatures $\vec{d}$.  This is the reason increasing disorder corresponds to increasing $\beta$.

The $\tau\left|d_i - d_{i+1}\right|$ term is an attempt to capture the influence of triple points, which spreads out curvature over adjacent triples of bands.  The term enforces some local correlation between the $d_i$ and $d_{i+1}$.

The appearance of absolute values may look strange but this is the correct scaling for energy/Chern number.  To see this, consider a composition
\[\arraycolsep=1.4pt\xymatrix{
T^2 \ar[r]^-{\text{\tiny $\left(\begin{array}{cc} n & 0 \\ 0 & n \end{array}\right)$}} \ar@/_20pt/[rr]^-g & T^2 \ar[r]^-f & BU(1)
}\]
$g^\ast c_1\left[T^2\right] = n^2 f^\ast c_1\left[T^2\right]$ and $\int_{T^2} d\,\text{area}\|\nabla g\|^2 = n^2\int_{T^2} d\,\text{area}\|\nabla f\|$ as well.

There is an invisible factor $n^{2 - 2}$ in the definition of $E$.  Passing from a band on $T_\text{big}^2$ to a similar band on the smaller $T_\text{small}^2$, the integrand of energy scales as $n^2$ but the integration measure as $n^{-2}$.

Given $E$, define the clearly convergent partition function: $Z(\beta) = \sum_{\vec{c}} e^{-\beta E\left(\vec{c}\right)}$, where the constraint on $\vec{c}$ is $\sum\vec{c} = c_1(e)$.  Then the model probability for observing a vector $\vec{c}$ of Chern numbers for the $n^2$ bonds over $T_\text{small}^2$ is:

\[p\!\left(\vec{c}\right) = \frac{1}{Z} e^{-\beta E\left(\vec{c}\right)}.\]

As explained above, $\beta \sim$ disorder strength, as at large disorder the distribution ceases to fluctuate corresponding to low temperature.  For a wildly fluctuating band $b$, $\tau$~should be larger since there will be more triple points of $\tilde{e}$, but for ordinary band geometries, $\tau$ may be fairly universal.


\bibliography{../bigbib.bib}

\end{document}